\documentclass[11pt]{article}

\usepackage{amssymb,amsmath}
\usepackage[totalwidth=17cm,totalheight=24cm]{geometry}
\usepackage{graphicx}
\usepackage{color}

\newcommand{\C}{{\mathbb C}}
\newcommand{\R}{{\mathbb R}}

\newcommand{\be}{\begin{eqnarray}}
\newcommand{\ee}{\end{eqnarray}}
\newcommand{\ba}{\nopagebreak[3]\begin{eqnarray}}
\newcommand{\ea}{\end{eqnarray}}
\newcommand{\id}{{\mathbb I}}

\begin{document}

\title{Isometries from gauge transformations}
\author{Ernesto Frodden${}^\star$ and Kirill Krasnov${}^\dagger$\\ {}\\
{\it ${}^\star$ Instituto de Sistemas Complejos de Valparaiso, Subida Artilleria 470, Valparaiso, Chile}\\ {}\\
{\it ${}^\dagger$ School of Mathematical Sciences, University of Nottingham, NG7 2RD, UK}}

\date{February 2020}
\maketitle
\begin{abstract}In four dimensions one can use the chiral part of the spin connection as the main object that encodes geometry. The metric is then recovered algebraically from the curvature of this connection. We address the question of how isometries can be identified in this ``pure connection" formalism. We show that isometries are recovered from gauge transformation parameters satisfying the requirement that the Lie derivative of the connection along a vector field generating an isometry is a gauge transformation. This requirement can be rewritten as a first order differential equation involving the gauge transformation parameter only. Once a gauge transformation satisfying this equation is found, the isometry generating vector field is recovered algebraically. We work out examples of the new formalism being used to determine isometries, and also prove a general statement: a negative definite connection on a compact manifold does not have symmetries. This is the precise ``pure connection" analog of the well-known Riemannian geometry statement that there are no Killing vector fields on compact manifolds with negative Ricci curvature. 
\end{abstract}

\section{Introduction}

A new formalism for Euclidean or Lorentzian four-dimensional geometry, which encodes geometry in the chiral part of the spin connection rather than the metric, has been proposed in \cite{Joel} and \cite{Krasnov:2011pp}, with the latter reference explaining how Einstein metrics can be treated. An exposition oriented at mathematicians appeared in \cite{Fine:2013qta}. In this formalism, Einstein equations are equivalent to a set of second-order partial differential equations on an ${\rm SO}(3)$ connection (${\rm SL}(2,\C)$ connection in the case of the Lorentzian signature). The metric is recovered algebraically from the curvature of this connection. Moreover, the equations on the connection follow by extermising a functional with some attractive convexity properties. There are no such convexity properties for the Einstein-Hilbert functional of the metric formalism. Thus, the connection formalism re-expresses the hard to solve Einstein equations as equations with different mathematical properties for a different object - the connection. It can thus be expected that some questions that are difficult in one formalism will be easier in the other. The first results confirming that this expectation is correct are the references \cite{Fine:2019tas} and \cite{Fine:2015hef}. The first of these uses the convexity of the pure connection action functional to prove a new, stronger than the one previously available, result on the local rigidity of 4D Einstein metrics. The second paper uses the connection formalism to describe the asymptotically hyperbolic case, and in particular gives a new, simpler prescription for the renormalised volume. 

The aim of this paper is to address the question of how isometries can be identified in this chiral pure connection formalism. Concretely, the question that is of interest to us is, given a connection, find the vector fields, if any, such that the Lie derivative of the metric defined by the connection in the direction of these vector fields is zero. Of course, one can answer this question by first computing the metric and then writing down the Killing equation. But this would be unduly complicated, and the aim of this paper is to develop a procedure that works directly at the level of the connection. For simplicity we do everywhere in the case of the Euclidean signature, but Lorentzian treatment is also possible. 

The Euclidean signature setup is that we have an ${\rm SO}(3)$ connection $A^i, i=1,2,3$. Let $F^i=dA^i + (1/2)\epsilon^{ijk} A^j A^k$  be the curvature 2-form. When $F^i\wedge F^j$ is a matrix-valued top form with a definite matrix (i.e. all eigenvalues are of the same sign) then the connection defines a Euclidean metric $g_A$ that is algebraically constructed from the curvature. We are interested in the question of whether this metric has Killing vector fields, and how to find them. 

The basic idea is simple. We take the Lie derivative of $A^i$ with respect to the would-be Killing vector field, and require that the result is a gauge transformation:
\be\label{main-equation}
{\cal L}_v A^i = d_A \xi^i.
\ee
This immediately implies that the metric constructed from the curvatures $F^i$ has vanishing Lie derivative with respect to $v$
\be
{\cal L}_v g_A = 0,
\ee
simply because the metric is a gauge-invariant construct. We don't even need to specify an explicit formula for how the metric is produced, it is sufficient to know that it produced in an algebraic fashion from $F^i$ that transform covariantly under gauge transformations, and the metric itself is gauge invariant. 

Thus, the question we would like to address is when the equation (\ref{main-equation}) has solutions. In this paper we describe some examples of solving the equation (\ref{main-equation}), as well as prove one general statement. The examples we consider are those of the hyperbolic space $H^4$, the 4-sphere $S^4$, as well as a more complicated example of a ``spherically-symmetric" connection. 

Our other goal is to establish a general statement as to existence of solutions of (\ref{main-equation}) on compact manifolds. To describe it, 
we need to introduce some additional notions. A connection is called {\bf definite} when the $3\times 3$ matrix appearing in $F^i\wedge F^j$ is definite at all points of $M$. Such connections have some attractive properties. Below we will explain that a definite connection has a well-defined notion of {\bf sign}. When the connection in question is Einstein, i.e., the metric it defines is Einstein, this sign correlates with the sign of the Ricci curvature. Thus, for instance, we will see below that the chiral connection on the four-sphere $S^4$ has the positive sign, while that on the hyperbolic space $H^4$ is negative. The statement that we prove is that a negative-definite connection on a compact manifold does not have symmetries. 

Our statement should be compared to the well-known Riemannian geometry statement that a negatively Ricci curved Riemannian metric has no isometries. The proof is simple, and we remind it to the reader for completeness. We have
\be
\int_M (\nabla_a v_b)^2 = -\int_M v^b \nabla^a \nabla_a v_b = \int_M v^b \nabla^a \nabla_b v_a.
\ee
Here, to get the first equality we integrated by parts, and so used the compactness of the four-manifold $M$. To get the second equality we used the Killing equation $\nabla_{(a} v_{b)}=0$. We can then use $\nabla_a v^a=0$, which is the contracted Killing equation to write
\be
\int_M (\nabla_a v_b)^2 = \int_M v^b (\nabla_a \nabla_b - \nabla_b \nabla_a) v^a = \int_M v^b R^a{}_{cab} v^c = \int_M R_{bc} v^b v^c.
\ee
When the Ricci tensor $R_{ab}$ is negative definite we have a non-negative right hand-side equal to a non-positive left hand-side, which implies that $v^a=0$. The statement that we prove is a precise analog of this Riemannian geometry statement in the land of connections. We also prove it by an integration by parts argument. 

First examples of negative-definite connection on compact four-manifolds that are not gravitational instantons (i.e. not anti-self-dual Einstein) appear in \cite{FP1}, while the first non-trivial examples of Einstein negative-definite connections are constructed in \cite{Fine-Premoselli}. Another relevant work is \cite{FP2}, where it is shown that a negative-definite connection on a compact four-manifold can never admit an $S^1$ symmetry. The methods used in this reference are completely different from ours here, but the result of \cite{FP2} can be used to give an alternative proof of our statement.\footnote{KK is grateful to Joel Fine for this argument.} Indeed, the group of symmetries of the definite connection is a compact Lie group G (just as the group of isometries of a compact Riemannian manifold is a compact Lie group). Then the flow of the given non-trivial symmetry generates a closed subgroup of G. This subgroup must be commutative (since it is the closure of the flow of a single vector field) and so is a copy of a torus inside G. It then follows that each non-trivial symmetry vector field gives rise to a circle action, which is impossible \cite{FP2}. 

There is also a relation between our work and the classical proof by N. Hitchin of the statement that the only positively curved gravitational instantons (i.e. anti-self-dual (ASD) Einstein metrics) are those on $S^4$ and $\C P^2$, see e.g. \cite{Besse}, Theorem 13.30. In fact, what Hitchin uses is precisely the ``projected" version of our main equation (\ref{main-equation}), see (\ref{eqn-proj}) below, in the situation when the matrix $X^{ij}$ appearing as the $3\times 3$ matrix in $F^i\wedge F^j$ is the identity matrix. This is the situation relevant for the gravitational instantons. Hitchin can then estimate the dimension of the space of solutions of this projected equation from the index of a certain complex of differential operators that becomes available on an instanton background. All considerations in the Hitchin's proof are in the connection territory rather than the metric, which again serves to illustrate the power of the chiral connection formalism. Our work can thus be read as that generalising (some of the) methods used in the Hitchin's proof to the case of chiral connections on general four-manifolds rather than ASD Einstein. 

The organisation of the rest of this paper is as follows. In next section, we remind the reader the basics of the chiral pure connection formalism. The notions of a definite connection and its sign are explained here. The Section \ref{sec:gen-stat} introduces the projected version of the equation (\ref{main-equation}), which is a set of first-order differential equations on a gauge parameter, and explains how isometries can be recovered from solutions of these. It also proves the main statement of this paper about negative-definite connections. Section \ref{sec:examples} describes a set of examples that are treated via our formalism. 

\section{Preliminaries: Definite connections}
\label{sec:prelim}

\subsection{Definite connections}

We first describe how metrics arise from ${\rm SO}(3)$ connections. Let $E$ be a rank 3 vector bundle over a 4-dimensional manifold $M$. We assume that $E$ is equipped with a positive definite metric in the fibres. We can use this metric to identify $E$ with its dual space $E^*$, so we will not make a distinction between these two spaces in what follows. We choose an orthonormal basis in $E$, so that the metric in the fibres has components $\delta^{ij}, i,j=1,2,3$. We also choose an orientation of $E$. In a basis it is described as an anti-symmetric  tensor $\epsilon^{ijk}$. We work with basis vectors oriented so as to give this tensor its standard components. 

Let $A$ be an ${\rm SO}(3)$ connection in $E$. Concretely, we think of $A$ in terms of its connection components $A^i, i=1,2,3$, which is a vector-valued one-form. Let $F^i = dA^i + (1/2)\epsilon^{ijk} A^j A^k$ be the curvature 2-form. 

We assume $M$ to be orientable. Let $\mu$ be a nowhere vanishing 4-form on $M$. A choice of $\mu$ defines the following $E\otimes E$ valued object $X^{ij}$ constructed from the curvature
\be\label{X}
F^i F^j = - 2 X^{ij} \mu.
\ee
The sign and the numerical factor in this formula are for future convenience. There is of course ambiguity in the choice of $\mu$, and different choices of $\mu$ are related by multiplication by a nowhere vanishing function.  If we change $\mu\to \Omega^4 \mu$ then $X^{ij}$ transforms $X\to \Omega^{-4}X$. It will be useful to track the effects of these transformations in what follows, as meaningful geometric objects will be invariant under these transformations. 

A connection $A$ is called {\bf definite} if the symmetric matrix $X^{ij}$ is definite at every point of $M$, i.e. has eigenvalues that are all of the same sign. A definite connection defines an orientation of $M$. This is an orientation $\mu$ in which $X^{ij}$ is positive definite. 

\subsection{Metrics from definite connections}

A definite connection equips $M$ with a unique conformal class of a Riemannian signature metric. This is obtained as follows. We define the metric pairing of two vectors $v,u\in TM$ as proportional to 
\be\label{Urb}
g(v,u) \mu \sim \sigma \epsilon^{ijk} i_u F^i i_v F^j F^k.
\ee
The proportionality sign here stands for modulo multiplication by a positive function on $M$, $\mu$ is the orientation 4-form that makes the matrix $X^{ij}$ defined in (\ref{X}) positive definite, and $\sigma$ is a sign that makes $g$ to be a Riemannian metric of signature all plus. When $\sigma>0$ ($\sigma<0$) the connection is called {\bf positive} ({\bf negative})  definite. The geometric meaning of this choice of the metric is that it is precisely the conformal metric that makes the triple of 2-forms $F^i$ {\bf self-dual} (or anti-self-dual, this depends on the choice of orientation, to be discussed below). 

There is also a preferred choice of the conformal factor for the above formula. Thus, a definite connection equips $M$ with the canonical metric defined as follows
\be\label{Urb-fixed}
g(v,u) \mu_g =   \frac{\sigma}{6} \epsilon^{ijk} i_u F^i i_v F^j F^k.
\ee
Here $\mu_g$ is the volume form for the metric $g$, in the orientation that makes $X^{ij}$ positive definite. This formula defines the metric uniquely, as well as specifies the sign $\sigma$ of the connection. 

We note, however, that there are in general other choices of the conformal class that can be made. The choice described is the one that is most mathematically natural, but other choice are possible, and in fact more appropriate in some situations. We will return to this point below after we describe a convenient formalism for doing calculations with definite connections.

\subsection{Basis in the space of self-dual 2-forms. Projectors}

In this subsection we will assume that a metric is given. We will go back to metrics from connections in the following subsection. Let $g_{\mu\nu}$ be a Euclidean signature metric tensor, and let $e^I, I=1,2,3,4$ be a co-frame for the metric $ds^2= g_{\mu\nu} dx^\mu dx^\nu = \sum_I e^I e^I$. We introduce the following basis in the space of self-dual 2-forms
\be\label{Sigma}
\Sigma^i = e^4 e^i - \frac{1}{2} \epsilon^{ijk} e^j e^k.
\ee
The wedge product of 1-forms here is omitted for compactness of notation. Note that $\Sigma^i\in E\otimes\Lambda^2$, i.e. these are vector-valued 2-forms. They take values in the vector bundle $E$ that has been introduced above. While it may not be obvious that these two vector bundles should be the same, it can be mentioned already at this stage that given the objects $\Sigma^i$ that come from a metric there is a unique ${\rm SO}(3)$ connection $A^i$ on $E$ that satisfies $d^A \Sigma^i=d\Sigma^i + \epsilon^{ijk}A^j \Sigma^k=0$. This explains why both the connection and the objects $\Sigma^i$ should be thought of as valued in the same ${\rm SO}(3)$ bundle $E$.

The 2-forms (\ref{Sigma}) are self-dual in the orientation $1234$, and we will refer to them as the basis of self-dual 2-forms for the metric given. Alternatively, they are anti-self-dual in the orientation $4123$ that is perhaps more natural with our conventions. So, calling them self- or anti-self-dual is a matter of taste and we use self-dual because it's shorter. 

One can raise one of the two indices of these 2-forms with the metric and obtain objects $\Sigma^i_\mu{}^\nu\in E\otimes{\rm End}(T^*M)$ that have the following algebra
\be\label{algebra}
\Sigma^i_\mu{}^\rho \Sigma^j_\rho{}^\nu = - \delta^{ij} \delta_\mu{}^\nu + \epsilon^{ijk} \Sigma^k_\mu{}^\nu.
\ee
Thus, the objects $\Sigma^i_\mu{}^\nu$ are 3 endomorphisms of the space of 1-forms that satisfy the algebra of imaginary quaternions. 

\subsection{Projectors}

Let us now consider the space $E\otimes \Lambda^1$ of vector-valued 1-forms $\eta_\mu^i$, where again $E$ is the space that the vectors with index $i$ take values in. The endomorphisms $\Sigma^i_\mu{}^\nu$ can be used to build the following operator acting on $E\otimes \Lambda^1$:
\be
J_\Sigma : E\otimes \Lambda^1 \to E\otimes \Lambda^1, \qquad (J_\Sigma \eta)^i_\mu := \epsilon^{ijk} \Sigma^j_\mu{}^\nu \eta^k_\nu.
\ee
It is not hard to check that the square of $J_\Sigma$ satisfies
\be
J_\Sigma^2 = 2 \id + J_\Sigma,
\ee
where $\id$ is the identity operator. This means that $J_\Sigma$ can be used to build the following two (orthogonal) projector operators
\be\label{projectors}
P^{(1,1)} := \frac{1}{3}(\id + J_\Sigma), \qquad P^{(3,1)} := \frac{1}{3}(2\id - J_\Sigma).
\ee
It is also not hard to see that $P^{(1,1)}$ is the projector on the elements in $E\otimes \Lambda^1$ of the form $\Sigma^i_\mu{}^\nu v_\nu$, where $v_\mu$ is an arbitrary 1-form. Such elements are then in the kernel of the projector $P^{(3,1)}$.

\subsection{Parametrisation of the curvature}

As discussed after (\ref{Urb}), there always exists a choice of the conformal metric that makes the triple of 2-forms $F^i$ self-dual. Choosing any representative in the conformal class of this metric, and then choosing a co-frame for it gives us the basis (\ref{Sigma}) in the space of self-dual 2-forms. Because $F^i$ are by construction self-dual, we can expand them into the basis of 2-forms $\Sigma^i$. We get
\be\label{F-Sigma}
F^i = \sigma \sqrt{X}^{ij} \Sigma^j.
\ee
This is a very important formula for the applications that follow, so let us explain all of its ingredients. The quantity $\sigma=\pm 1$ is the sign of the connection that was already defined in (\ref{Urb}). The object $\sqrt{X}^{ij}$ is the positive matrix square root of positive definite matrix $X^{ij}$ that was defined in (\ref{X}). The definition of $X^{ij}$ depends on a choice of the volume form $\mu$, and so is the choice of a metric in the conformal class of metrics defined by (\ref{Urb}). The objects $\Sigma^i$ are thus also dependent on a choice of the metric. It is natural to relate these two ambiguities. This is best done by setting
\be\label{Sigma-Sigma}
\Sigma^i \Sigma^j = - 2\delta^{ij} \mu,
\ee
thus setting the volume form $\mu$ to be the metric volume form. There is also a choice of orientation that has been made in writing this formula, it is clear that we prefer the $4123$ orientation that makes $\Sigma^i$ anti-self-dual. With these choices the objects $X^{ij}$ and $\Sigma^i$ still depend on a choice of a representative in the conformal class of (\ref{Urb}). One passes between different choices by the transformations $\Sigma^i\to \Omega^2 \Sigma^i$ and $\sqrt{X}^{ij} \to \Omega^{-2}  \sqrt{X}^{ij} $ so that $F^i$ is independent of such choices. 

Finally, it is clear that the matrix of coefficients appearing in the decomposition of $F^i$ into the basis of $\Sigma^i$ must be $\sqrt{X}^{ij}$ because we must have (\ref{X}), which indeed results in view of (\ref{Sigma-Sigma}). 

The formula (\ref{F-Sigma}) provides a very convenient parametrisation of the curvature. The latter is parametrised by the metric it defines via (\ref{Urb}), as well as by the matrix $X^{ij}$ it defines via (\ref{X}). 

\subsection{Choice of a metric in the conformal class}

We now return to the question of how a metric in the conformal class of (\ref{Urb}) can be fixed. It is clear that fixing a metric also fixes the volume form $\mu$. But this turns out to be not the best way of eliminating the conformal freedom present so far. 

To develop a better way of fixing this freedom let us discuss how this freedom is eliminated in the formula (\ref{Urb-fixed}) that specifies the metric completely. To this end, we substitute the parametrisation (\ref{F-Sigma}) into (\ref{Urb-fixed}). This gives
\be
g(u,v) \mu_g = \frac{1}{6} {\rm det}(\sqrt{X}) \epsilon^{ijk} i_u \Sigma^i i_v \Sigma^j \Sigma^k.
\ee
On the other hand, a direct calculation with $\Sigma^i$ defined by (\ref{Sigma}) gives
\be
g(u,v) \mu_g = \frac{1}{6}  \epsilon^{ijk} i_u \Sigma^i i_v \Sigma^j \Sigma^k,
\ee
which implies that we have
\be\label{det-cond}
{\rm det}(\sqrt{X})=1.
\ee
This means that in the case of metric fixed by (\ref{Urb-fixed}) we have fixed the metric in the conformal class of (\ref{Urb}) by imposing a single condition on the components of the matrix $X^{ij}$. Note that this is of course an ${\rm SO}(3)$-invariant condition. 

It is now clear that we can more generally fix a metric in the conformal class of (\ref{Urb}) by imposing a single ${\rm SO}(3)$-invariant condition on the matrix $X^{ij}$. 

\subsection{Example: choice relevant for imposing the Einstein condition}

It turns out that if one wants to impose the condition that some metric in the conformal class of (\ref{Urb}) is Einstein, the relevant metric is not the one given by (\ref{Urb-fixed}), but rather corresponding to a different condition on $X^{ij}$. Namely, the condition is ${\rm Tr}(\sqrt{X})=const$. This is explained in \cite{Krasnov:2011pp}, and in a more mathematician-friendly way in \cite{Fine:2013qta}. 

\section{Killing vectors of the metric defined by a definite connection}
\label{sec:gen-stat}

As we discussed in the previous section, a definite connection defines the conformal class of a metric via formula (\ref{Urb}). A metric in this conformal class can be singled out if one imposes a single ${\rm SO}(3)$-invariant condition on the matrix $X^{ij}$, of a type that fixes the freedom of conformal rescalings $X^{ij}\to \Omega^{-4} X^{ij}$. For example, this can be the condition (\ref{det-cond}), or the conditions ${\rm Tr}(\sqrt{X})=const$ that is relevant for imposing the Einstein condition on the metric. In what follows we will assume that such a choice is made, but it will not be important which of these options is selected. We will see that the main statement will be valid for any of the choices of the type we discussed. 

We now ask the following question. Given a definite connection on $M$, together with a choice of the metric in the conformal class of (\ref{Urb}), how can we find isometries of the metric just by looking at the connection? The answer has already been explained in the Introduction. Indeed, it is sufficient that the Lie derivative of the connection is a gauge transformation
\be\label{LA}
{\cal L}_v A^i = d^A \xi^i.
\ee
To see that a vector field that satisfies this condition is a Killing vector field of the metric constructed from the connection it is sufficient to note that the metric is constructed algebraically from the curvature of $A^i$. This is the case for both the conformal metric, constructed via (\ref{Urb}), as well as the conformal factor that is fixed by some ${\rm SO}(3)$-invariant condition on the matrix $X^{ij}$. Both of these are ${\rm SO}(3)$-invariant, and so the Lie derivative of the metric, which in view of (\ref{LA}) is a gauge transformation, vanishes. 

Thus, to find symmetries of a connection, as well as those of the metric it defines, we need to find all solutions of (\ref{LA}). This is an overdetermined first-order differential equation on both the vector field $v$ and the gauge transformation parameter $\xi^i$. In the following section we will describe examples of how this equation can be solved. The purpose of this section is to describe some general properties of solutions of this equation.

\subsection{Extracting an equation for the gauge parameter}

As we now discuss, to solve (\ref{LA}) it is best to rewrite it in a way that eliminates the vector field from this equation, resulting in a first order differential equation for the gauge parameter only. This, however, requires a redefinition of the gauge transformation parameter. 

The equation (\ref{LA}) can be rewritten as
\be\label{eqn-F}
i_v F^i = d^A \tilde{\xi}^i, 
\ee
where we introduced the notation
\be\label{shifted}
\tilde{\xi}^i := \xi^i - i_v A^i.
\ee
As we shall now see, the left-hand-side of the equation (\ref{eqn-F}) can be projected away, with the result being a differential equation for the gauge transformation parameter $\tilde{\xi}^i$ only. After this is found one can find the vector field $v$ algebraically from the covariant derivative of $\tilde{\xi}^i$. 

To obtain an equation for $\tilde{\xi}^i$ we use the parametrisation of the curvature (\ref{F-Sigma}). Multiplying this equation from both sides with $\sqrt{X}^{-1}$ we get
\be\label{eqn-Sigma}
\sigma \,i_v \Sigma^j = (X^{-1/2})^{ij} d^A \tilde{\xi}^j.
\ee
As we know from (\ref{projectors}), the left-hand-side of this equation is in the kernel of the projector $P^{(3,1)}$. Using the explicit form of this projector we get
\be\label{eqn-proj}
2 (X^{-1/2})^{ij} d_\mu^A \tilde{\xi}^j = \epsilon^{ijk} \Sigma_\mu^j{}^\nu (X^{-1/2})^{kl} d_\nu^A \tilde{\xi}^l.
\ee
This is the desired equation that only contains the gauge parameter.

\subsection{Determining the vector field once the gauge parameter is solved for}

Let us now describe how the vector field can be recovered once the equation (\ref{eqn-proj}) is solved. The way to do this can be seen from (\ref{eqn-Sigma}). Indeed, in index notations the left-hand-side reads $\sigma v^\nu \Sigma^i_{\nu\mu}$. Multiplying this with $\Sigma^i_\rho{}^\mu$, and using the algebra (\ref{algebra}) we have $3\sigma v_\rho$. This means that
\be\label{v}
v_\mu = \frac{\sigma}{3} \Sigma^i_\mu{}^\nu (X^{-1/2})^{ij} d_\nu^A \tilde{\xi}^j.
\ee
This means that when $\tilde{\xi}^i$ is found from (\ref{eqn-proj}), we can recover the vector field algebraically from the derivative of the gauge parameter. Thus, the problem reduces to that of solving (\ref{eqn-proj}).

One consequence of the equation (\ref{v}) is that 
\be\label{v2}
|v|^2 = \frac{1}{9} \Sigma^i_\mu{}^\nu (X^{-1/2})^{ij} d_\nu^A \tilde{\xi}^j \Sigma^{k\, \mu\rho} (X^{-1/2})^{kl} d_\rho^A \tilde{\xi}^l \\ \nonumber
=\frac{1}{9} ( \delta^{ik} g^{\nu\rho} - \epsilon^{iks}\Sigma^{s\, \nu\rho}) (X^{-1/2})^{ij}(X^{-1/2})^{kl} d_\nu^A \tilde{\xi}^j d_\rho^A \tilde{\xi}^l \\ \nonumber
=\frac{1}{9} (X^{-1})^{ij} g^{\mu\nu} d_\mu^A \tilde{\xi}^i d_\nu^A \tilde{\xi}^j - \frac{1}{9} {\rm det}(X^{-1/2}) \epsilon^{ijk} d^A_\mu \tilde{\xi}^i d^A_\nu \tilde{\xi}^j (X^{1/2})^{kl} \Sigma^{l\,\mu\nu}.
\ee
We will see that these two terms are actually multiples of each other. 

\subsection{Determining the gauge parameter from the vector field}

For completeness, let us note that one can also solve for $\tilde{\xi}^i$ algebraically if one knows the vector field. To see this, we take the exterior covariant derivative of the equation (\ref{eqn-F}). We have
\be
d_A (i_v F^i) = \epsilon^{ijk} F^j \tilde{\xi}^k.
\ee
This is an algebraic equation for $\tilde{\xi}^i$ that can be solved. Indeed, writing everything in index notation and replacing $F^i$ with its parametrisation (\ref{F-Sigma}) we have
\be
2 \left[d_A(\sqrt{X}^{ij}  i_v \Sigma^j)\right]_{\mu\nu} = \epsilon^{ijk} (\sqrt{X})^{jm} \Sigma^m_{\mu\nu} \tilde{\xi}^k.\nonumber
\ee

The gauge parameter can be extracted by multiplying this with $\epsilon^{ilp} (X^{-1/2})^{ln} \Sigma^{n\mu\nu}$. We get
\be\label{xi}
\tilde{\xi}^k=\frac{1}{4} \epsilon^{kli} (X^{-1/2})^{ln} \Sigma^{n\mu\nu} \left[d_A(\sqrt{X}^{ij}  i_v \Sigma^j)\right]_{\mu\nu}.
\ee

Thus, in principle, the gauge parameter can be computed in terms of the derivatives of the vector field.

\subsection{Rewriting the equation}

The most convenient form of the equation (\ref{eqn-proj}) is obtained by multiplying it on both sides with another factor of $X^{-1/2}$. We get
\be\label{eqn-rewr-1}
2 (X^{-1})^{ij} d_\mu^A \tilde{\xi}^j = {\rm det}(X^{-1/2}) \epsilon^{ijk} (X^{1/2})^{jl} \Sigma^l_\mu{}^\nu d^A_\nu \tilde{\xi}^k.
\ee
The reason for writing it in this form is that we can now use (\ref{F-Sigma}) to rewrite the right-hand-side in terms of the curvature
\be
2 (X^{-1})^{ij} d_\mu^A \tilde{\xi}^j = \sigma {\rm det}(X^{-1/2}) \epsilon^{ijk} F^j_\mu{}^\nu d^A_\nu \tilde{\xi}^k.
\ee
Note that the metric is still needed to write this equation, as one needs to raise one of the indices of the curvature on the right-hand-side. 

We can also use (\ref{eqn-rewr-1}) to further manipulate (\ref{v2}). Indeed, from (\ref{eqn-rewr-1}) we have see that the second term in the last line of (\ref{v2}) is twice the first term. This means that we can write (\ref{v2}) in two different ways as
\be\label{v2-1}
|v|^2= \frac{1}{3} (X^{-1})^{ij} g^{\mu\nu} d_\mu^A \tilde{\xi}^i d_\nu^A \tilde{\xi}^j = -\frac{1}{6} {\rm det}(X^{-1/2}) \epsilon^{ijk} d^A_\mu \tilde{\xi}^i d^A_\nu \tilde{\xi}^j (X^{1/2})^{kl} \Sigma^{l\,\mu\nu}.
\ee

\subsection{The case of a compact manifold}

We now deduce some consequence of (\ref{v2-1}) on a compact manifold. We multiply the second equation in (\ref{v2-1}) by ${\rm det}(\sqrt{X})$ and get integrate over $M$. We get
\be
6 \int_M {\rm det}(\sqrt{X}) |v|^2 \mu_g = - \int_M \epsilon^{ijk} d^A_\mu \tilde{\xi}^i d^A_\nu \tilde{\xi}^j (X^{1/2})^{kl} \Sigma^{l\,\mu\nu} \mu_g.
\ee
We remind that $\mu_g$ is the volume form of the metric for which $\Sigma^i$ are the basis of self-dual 2-forms, and with respect to which $|v|^2$ is evaluated. The key point now is that we can rewrite the integrand on the right-hand-side in terms of the curvature using (\ref{F-Sigma}), and then in terms of the wedge product of forms using the (anti-)self-duality of the curvature. We get
\be
6 \int_M {\rm det}(\sqrt{X}) |v|^2 \mu_g =  \sigma \int_M \epsilon^{ijk} d^A \tilde{\xi}^i d^A \tilde{\xi}^j F^k,
\ee
where now the right-hand-side contains the wedge product of forms. We then integrate by parts and use the Bianchi identity for the curvature to get
\be
6 \int_M {\rm det}(\sqrt{X}) |v|^2 \mu_g = -\sigma \int_M  \epsilon^{ijk} \tilde{\xi}^i \epsilon^{jmn} F^m \tilde{\xi}^n F^k.
\ee
Using the definition (\ref{X}) we get, finally
\be\label{compact}
3 \int_M {\rm det}(\sqrt{X}) |v|^2 \mu_g = \sigma \int_M  ({\rm Tr}(X) \delta^{ij} - X^{ij}) \tilde{\xi}^i\tilde{\xi}^j\, \mu_g.
\ee
By construction the matrix $X^{ij}$ is positive definite, and so the left-hand-side is non-negative. The integrand on the right-hand-side is also non-negative because ${\rm Tr}(X)\delta^{ij}-X^{ij}$ is also positive-definite. This means that for negative definite connections $\sigma<0$ both sides must vanish, and there are no symmetries. We learn that negative definite connections on compact manifolds cannot have symmetries. This is a precise analog of the classical result in Riemannian geometry that compact manifolds with negative Ricci curvature cannot have symmetries. 

Of course the above discussion relies on an integration by parts argument and is only true on compact manifolds. There definitely are symmetries on the hyperbolic space $H^4$, as our next section example demonstrates. 

\section{Examples}
\label{sec:examples}

We now consider several examples of how the procedure described above can be put to use in practice. We start with the simplest example of the hyperbolic space, then consider the four-sphere, and finally consider the more involved example of a ``spherically-symmetric" connection. 

\subsection{Hyperbolic space}

We take a connection on the upper-half-space $t>0$ in $\R^4$ with coordinates $t,x^i$. The connection in question is 
\be\label{A}
A^i = \frac{1}{t} dx^i,
\ee
and is the self-dual part of the Levi-Civita connection for the hyperbolic metric
\be\label{metric}
ds^2 = \frac{1}{t^2}( dt^2 + \sum_i (dx^i)^2).
\ee
This metric arises from the connection as follows. The curvature of the connection (\ref{A}) is given by
\be\label{F-Sigma-H4}
F^i = -\Sigma^i, \qquad \Sigma^i = \frac{1}{t^2} (dt dx^i - \frac{1}{2}\epsilon^{ijk} dx^j dx^k).
\ee
One first obtains the conformal metric as the unique conformal metric that makes the triple of curvatures self-dual. One then fixes the conformal factor so that the volume form is the one appearing on the right-hand-side of 
\be 
F^i F^j = - 2\delta^{ij} v.
\ee
This gives the metric (\ref{metric}).

Let us now search for Killing vector fields. We write a general vector field as
\be
v = v^t \frac{\partial}{\partial t} + v^i \frac{\partial}{\partial x^i},
\ee
where $v^t, v^i$ are functions of all the coordinates. The Lie derivative of $A^i$ is then
\be
{\cal L}_v = -\frac{v^t}{t^2} dx^i + \frac{1}{t} dv^i = \frac{1}{t} \frac{\partial v^i}{\partial t} dt + \left( \frac{1}{t} \frac{\partial v^i}{\partial x^j} - \frac{v^t}{t^2} \delta^{ij}\right) dx^j.
\ee
We want this to be equal to the gauge transformation with a parameter $\xi^i$, i.e. 
\be
d_A \xi^i = \frac{\partial \xi^i}{\partial t} dt + \left( \frac{\partial \xi^i}{\partial x^j} + \frac{1}{t} \epsilon^{ijk} \xi^k\right) dx^j,
\ee
where we have used the explicit expression (\ref{A}) for the connection. Matching the terms gives the following equations
\be\label{eqs}
\frac{\partial \xi^i}{\partial t} = \frac{1}{t} \frac{\partial v^i}{\partial t}, \qquad \frac{\partial \xi^i}{\partial x^j} + \frac{1}{t} \epsilon^{ijk} \xi^k= \frac{1}{t} \frac{\partial v^i}{\partial x^j} - \frac{v^t}{t^2} \delta^{ij}.
\ee

We note that differentiating the second equation with respect to time and using the derivative of the first equation with respect to the $x^j$ coordinates, these two equations imply
$$\epsilon^{ijk} \xi^k = \epsilon^{ijk} \frac{\partial v^k}{\partial t} + \frac{\partial v^i}{\partial x^j} + t^2 \frac{\partial}{\partial t} \left( \frac{v^t}{t^2}\right) \delta^{ij},$$
which gives us $\xi^i$ if we know the Killing vector field components $v^t, v^i$. 

\subsubsection{Solutions with $\xi^i=0$}

We now attempt to solve for $\xi^i$ and $v^t,v^i$ simultaneously. The easiest case is when $\xi^i=0$. We see that the vector field leaves the connection unchanged when 
\be
\frac{\partial v^i}{\partial t}=0, \qquad  \frac{\partial v^i}{\partial x^j} =\frac{v^t}{t} \delta^{ij}.
\ee
Thus, the first equation says that $v^i$ is time coordinate independent. In order for the left-hand-side of the second equation to be time independent we must have $v^t\sim t$. There are two possible solutions:

\noindent{\bf Case 1:} $v^t=0$, and $v^i$ is a constant vector. These are the 3 Killing vector fields corresponding to translations in the $x^i$ directions
$$v = v^i \frac{\partial}{\partial x^i}.$$

\noindent{\bf Case 2:} $v^t=t$, and $v^i =x^i$. This is the dilatations Killing vector field
$$v = t\frac{\partial}{\partial t} + x^i \frac{\partial}{\partial x^i}.$$

\subsubsection{Solutions with $\xi^i\not=0$}

Having found all vector fields that simply leave the connection invariant, let us find those whose effect can be offset by a gauge transformation.

\noindent{\bf Case 1:} The easiest case to consider is when $\xi^i$ is time independent. In this case $v^i$ is also time independent. The terms in the second equation in (\ref{eqs}) then have different powers of $1/t$ in front of them. It is only possible to satisfy this equation if $v^t=0$ and $\xi^i$ is a constant vector $\xi^i=\alpha^i$. This gives
$$ v= \epsilon^{ijk} x^j \alpha^k  \frac{\partial}{\partial x^i},$$
which is the vector field describing a rotation around the $\alpha^i$ axis. There is also an integration constant in obtaining this relation, but this integration constant is the already described Killing vector field corresponding to translations.

\noindent{\bf Case 2:} The only solution not covered is that corresponding to special conformal transformations. We simply state it, and develop tools to find it systematically later. We have
\be\label{special-conf}
 v^t=-2t \alpha^i x^i, \quad v^i = (t^2+|x|^2) \alpha^i - 2x^i (\alpha^j x^j), \quad \xi^i = 2t \alpha^i + 2\epsilon^{ijk} \alpha^j x^k,
 \ee
where $\alpha^i$ is a constant vector. 

\subsubsection{Systematic procedure: projection}
\label{systematic}

We now search for isometries using the systematic procedure described in Section \ref{sec:gen-stat}. The main equation (\ref{main-equation}) is rewritten as
\be
i_v F^i = d_A \tilde{\xi}^i, \qquad
\tilde{\xi}^i := \xi^i - i_v A^i.
\ee
As before, $i_v$ is the operator of the interior product, inserting the vector field $v$ into a differential form that follows. Given that the matrix $X^{ij}$ in question is the identity matrix, the projected equation (\ref{eqn-proj}) reads 
\be\label{proj-eqn-H4}
2 d^A_\mu \tilde{\xi}^i - \epsilon^{ijk} \Sigma^j_\mu{}^\nu d^A_\nu \tilde{\xi}^k=0.
\ee
For our background (\ref{F-Sigma-H4}) we have
\be
\Sigma^i_\mu{}^\nu = (dt)_\mu \left( \frac{\partial}{\partial x^i}\right)^\nu - (dx^i)_\mu \left( \frac{\partial}{\partial t}\right)^\nu - \epsilon^{ijk} (dx^j)_\mu \left( \frac{\partial}{\partial x^k}\right)^\nu,
\ee
and (\ref{proj-eqn-H4}) reduces to the following two equations
\be
2 \frac{\partial \tilde{\xi}^i}{\partial t} = \epsilon^{ijk} \frac{\partial \tilde{\xi}^k}{\partial x^j} - \frac{2}{t} \tilde{\xi}^i, \\ \nonumber
2\frac{\partial \tilde{\xi}^i}{\partial x^j} + \frac{1}{t} \epsilon^{ijk} \tilde{\xi}^k = \delta^{ij}\frac{\partial \tilde{\xi}^k}{\partial x^k} - \frac{\partial \tilde{\xi}^j}{\partial x^i}-\epsilon^{ijk} \frac{\partial \tilde{\xi}^k}{\partial t}.
\ee
The second equation splits into its symmetric and anti-symmetric with respect to $ij$ parts. The anti-symmetric part coincides with the first equation, while the symmetric part gives
\be\label{symm-part}
\frac{\partial \tilde{\xi}^i}{\partial x^j} + \frac{\partial \tilde{\xi}^j}{\partial x^i} = \frac{2}{3}\delta^{ij} \frac{\partial \tilde{\xi}^k}{\partial x^k} ,
\ee
which just says that the symmetric part of the spatial derivative of $\tilde{\xi}^i$ with respect to $x^j$ is proportional to $\delta^{ij}$. 
We rewrite the first equation in the following suggestive form
\be\label{eqn-xi}
D(t\tilde{\xi}) =0, 
\ee
where $D$ is a first order operator that maps vectors into vectors
\be
(D\lambda)^i = \left( \delta^{ij} \frac{\partial}{\partial t} + \frac{1}{2} \epsilon^{ijk} \frac{\partial}{\partial x^k} \right) \lambda^j.
\ee
We are thus interested in vectors in the kernel of the Dirac-like operator $D$, satisfying in addition (\ref{symm-part}).

\subsubsection{Finding the vector field}
\label{finding}

Let us assume that a solution for $\tilde{\xi}^i$ is found. The vector field can then be extracted by simple operations. Thus, we have for the background at hand
\be
- v^\nu \Sigma^i_{\nu\mu} = d^A_\mu \tilde{\xi}^i.
\ee
It follows that
\be
v^{\mu} = - \frac{1}{3} \Sigma^{i\,\mu\nu} d^A_\nu \tilde{\xi}^i.
\ee
For the case at hand this results in the following expressions
\be
v^t = \frac{t^2}{3} \frac{\partial \tilde{\xi}^k}{\partial x^k}, \qquad v^i = \frac{t^2}{3} \left( \frac{\partial \tilde{\xi}^i}{\partial t} - \frac{2}{t} \tilde{\xi}^i - \epsilon^{ijk} \frac{\partial \tilde{\xi}^j}{\partial x^k}\right).
\ee
The second expression can be rewritten as
\be
v = \left( t\frac{\partial}{\partial t} - 1\right) (t\tilde{\xi}) - \frac{2t}{3} D (t\tilde{\xi}),
\ee
where we used index-free notation. On solutions to (\ref{eqn-xi}) the second term drops and the spatial component of the vector field is simply recovered from the time derivative of $t\tilde{\xi}^i$.

\subsubsection{Solution}
\label{solution}

We now search for solutions of (\ref{eqn-xi}). The easy solution is any constant vector $t\tilde{\xi}^i =\lambda^i$. This gives the vector field corresponding to spatial translations. Let us search for more general solutions among expressions homogeneous in $t,x^i$ coordinates. At homogeneity degree one we can make the following ansatz
\be
t\tilde{\xi}^i = t \alpha^i + \epsilon^{ijk} \beta^j x^k + \gamma x^i,
\ee
where $\alpha^i,\beta^i$ are constant vectors and $\gamma$ is a constant. It is clear that both equations (\ref{eqn-xi}) and (\ref{symm-part}) are satisfied for any $\gamma$, and so $t\tilde{\xi}^i=x^i$ is the solution that corresponds to dilatations. For the remainder we have $D(t\tilde{\xi})=0$ implying $\alpha+\beta=0$, and so we get another solution $t\tilde{\xi}^i = t\alpha^i - \epsilon^{ijk} \alpha^j x^k$. It automatically satisfies (\ref{symm-part}) and corresponds to rotations. 

We now search for solutions in an expression of homogeneity degree two in $t, x^i$. The most general ansatz is
\be
t\tilde{\xi}^i = (t^2 + b |x|^2) \alpha^i + t\epsilon^{ijk} \beta^j x^k + x^i( \gamma^j x^j),
\ee
where $\alpha^i,\beta^i,\gamma^i$ are constant vectors and $b$ is a constant. The equation (\ref{eqn-xi}) gives $ 2\alpha^i -\beta^i=0$ and $\beta^i + b\alpha^i - (1/2)\gamma^i=0$. The equation (\ref{symm-part}) gives $2b\alpha^i + \gamma^i=0$. Together they imply that $b=-1$ and give the following solution
\be
t\tilde{\xi}^i = (t^2 - |x|^2) \alpha^i + 2t\epsilon^{ijk}\alpha^j x^k + 2x^i (\alpha^j x^j),
\ee 
which is easily seen to correspond to special conformal transformations (\ref{special-conf}). 

\subsection{Four-sphere}

The description of symmetries for the four-sphere geometry $S^4$ is very similar to the one in the hyperbolic geometry. The reason why we perform this exercise is to do an explicit check that the formula (\ref{compact}) holds. We use the conformally flat parametrisation of the metric. Let $x^1,\ldots,x^4$ be the Cartesian coordinates on $\R^4$.

To define the connection consider the conformal function $\Omega=2/(1+|x|^2)$. The metric of $S^4$ is 
\be\label{metricS4}
ds^2=\Omega^2((dx^4)^2+\sum_i (dx^i)^2),
\ee
where we singled out the fourth coordinate, and the index $i=1,2,3$.
The basis of chiral 2-forms is given by 
\be\label{SigmaS4}
\Sigma^i=\Omega^2(dx^4dx^i-\frac{1}{2}\epsilon^{ijk}dx^jdx^k).
\ee
The chiral connection satisfying $d^A \Sigma^i=0$ is then checked to be
\be\label{connectionS4}
A^i_\mu=\Sigma^i_\mu{}^\nu\partial_\nu( \ln \Omega).
\ee
This is checked using the (anti-) self-duality of $\Sigma^i_{\mu\nu}$ as well as the algebra (\ref{algebra}). The matrix of wedge-products $F^iF^j$ of the curvature is positive definite when the orientation as in \eqref{X} is chosen $\mu\sim dx^4dx^1dx^2 dx^3$. Then $X^{ij}=\delta^{ij}$ and $F^i=\Sigma^i$, which corresponds to a positive definite connection.

\subsubsection{$S^4$ isometries}

To obtain the parameters $\tilde\xi$ we solve the equation \eqref{eqn-rewr-1}. All calculations are entirely analogous to the hyperbolic case, with the only difference being that the rescaled gauge transformation parameter $\bar\xi^i$ is now defined via $\tilde \xi^i=\Omega\, \bar\xi^i$. The rescaled gauge parameter satisfies
\be
\frac{\partial \bar{\xi}^i}{\partial x^j} + \frac{\partial \bar{\xi}^j}{\partial x^i} = \frac{2}{3}\delta^{ij} \frac{\partial \bar{\xi}^k}{\partial x^k} ,
\ee
and
\be
D(\bar \xi)=0.
\ee
The equations are equal to \eqref{symm-part} and \eqref{eqn-xi} but for the $\bar\xi^i$ variable instead. They were already solved in section \ref{solution}. Therefore, here we just list the ten Killing gauge transformation parameters
\ba\nonumber
\tilde\xi^i_\lambda&=&\Omega\, \lambda^i\\ \label{xi-s4}
\tilde\xi^i_\gamma&=&\gamma\,\Omega\,  x^i\\ \nonumber
\tilde\xi^i_\alpha&=&\Omega \left(x^4\alpha^i - \epsilon^{ijk} \alpha^j x^k\right)\\\nonumber
\tilde\xi^i_\beta&=&\Omega \left(((x^4)^2 - |\vec x|^2) \beta^i + 2x^4\epsilon^{ijk}\beta^j x^k + 2x^i (\beta^j x^j)\right),
\ea
 with $\lambda^i,\, \gamma, \alpha^i$ and $\beta^i$ being ten constants corresponding to the ten Killing symmetries of $S^4$.

\subsubsection{From Killing vector fields to gauge transformations}

With the gauge transformation parameters just found it is possible to recover the Killing vector fields from \eqref{v}. This exercise was already performed in \ref{finding} for the hyperbolic geometry case. 
Since both examples are very similar, we feel it is more instructive to do the opposite check. Given the usual Killing vector fields we check that the gauge transformation parameters obtained with \eqref{xi} are in correspondence with the ones found in (\ref{xi-s4}).

The $S^4$ metric is conformally flat and the six rotational Killing vectors field for flat spacetime are also Killing vectors of $S^4$.
The vector components are
\be
v_{[\omega]}^\mu=\omega^{\mu\nu}x^\nu, 
\ee 
where $\omega^{\mu\nu}$ is an arbitrary constant anti-symmetric matrix. Here all index contractions are carried out using the flat metric on $\R^4$. There are also the four "translation" Killing vector fields
\be
v_{[\rho]}^{\mu}= \rho^{\mu}\frac{1-|x|^2}{2}- x^\mu (\rho^{\nu} x^\nu).
\ee
For the rotations, the corresponding six Killing parameters are recovered from \eqref{xi}, and read
\be
\xi^i_{[\omega]}=\frac{1}{2}\left(\omega^{\alpha\mu}+\omega^{\alpha\nu}x^\nu \partial^\mu \ln \Omega \right)\tilde{\Sigma}^i_{\mu\alpha}.
\ee
Here $\tilde{\Sigma}^i = \Omega^{-2}\Sigma^i$ is the flat metric ASD 2-forms. This gauge transformation parameter is a combination of $\tilde\xi^i_{[\lambda]}$ and $\tilde\xi^i_{[\beta]}$ from (\ref{xi-s4}). For the ``translations'' the four Killing parameters are
\ba
\tilde \xi^i_{[\rho]}&=&\Omega\, (\rho^{4}x^i-\rho^{i}x^4-\epsilon^{ijk}\rho^j x^k)\\ \nonumber
&=&\Omega  \rho^\mu x^\nu \tilde{\Sigma}^i_{\mu\nu}.
\ea
This again can be written as a combination of $\tilde\xi^i_{[\gamma]}$ and $\tilde\xi^i_{[\alpha]}$ from (\ref{xi-s4}).

We can now perform an explicit check of the formula (\ref{compact}). In this case the matrix $X^{ij}=\delta^{ij}$, and the check reduces to the computation of the integrals of the norms $|v|^2$ and $|\tilde{\xi}|^2$. We have explicitly computed the integrals and checked that the formula (\ref{compact}) holds. 

\subsection{Spherically symmetric connection}

We also work out the case of a static and ``spherically-symmetric" connection, and see how the corresponding Killing vector fields can be recovered. The increased level of difficulty of this example is that the matrix $X^{ij}$ is no longer a multiple of the identity matrix. This examples gives a good illustration of the difficulties of a generic setup.

Let us take the following Ansatz for the connection
\be
A^1 = a dt + \cos\theta d\phi, \qquad A^2=-b\sin\theta d\phi, \qquad A^3=b d\theta.
\ee
Here $\theta,\phi$ are the usual spherical coordinates and $a,b$ are functions of a ``radial" coordinate that we will call $R$. We reserve the name $r$ for the coordinate that is related to the area of the spheres of symmetry. For future reference, the curvature 2-forms are
\ba
F^1&=& - a' dt dR + (b^2-1) \sin\theta\, d\theta d\phi, \\ \nonumber
F^2&=& - ab \, dt d\theta + b' \sin\theta\, d\phi dR, \\ \nonumber
F^3&=& - ab \sin\theta\, dt d\phi + b' dR d\theta.
\ea

From the \eqref{Urb-fixed} with $\sigma=-1$ and volume $\mu_g=\sqrt{\det(g)}\, dtdRd\theta d\phi$ we have
\be\label{metric-spherical}
ds^2=\frac{a^2b^2a'}{f(R)}\, dt^2+\frac{a'b'^2}{f(R)}\, dR^2+\frac{ab(b^2-1)b'}{f(R)}(d\theta^2+\sin^2\theta\, d\phi^2),
\ee
with $f(R)=\left(a^2b^2(b^2-1)a'b'^2\right)^{1/3}$.

From \eqref{X} we have
\be
X^{ij}=\frac{1}{f(R)}\text{diag}\left(a'(b^2-1),abb',abb'\right).
\ee
Note $\det(X)=1$.

The two-forms basis can be read from \eqref{F-Sigma}
\ba
\Sigma^1&=&\sqrt{\frac{f(R)}{{a'(b^2-1)}}}\left(a'\, dtdR-(b^2-1)\sin\theta\, d\theta d\phi\right),\\
\Sigma^2&=&\sqrt{\frac{f(R)}{abb'}}\left(ab\, dtd\theta-b'\sin\theta\, dRd\phi\right),\\
\Sigma^3&=&\sqrt{\frac{f(R)}{abb'}} \left(ab\sin\theta \, dtd\phi-b'\, dR d\theta\right).
\ea

The metric (\ref{metric-spherical}) has the following four Killing vector fields
\ba\label{KVF-spherical}
v_{[t]}&=&\frac{\partial}{\partial t}\quad ,\\
v_{[\phi]} &=&\frac{\partial}{\partial\phi}\quad ,\\
v_{[1]}&=&\cos\phi\, \frac{\partial}{\partial\theta}-\cot\theta\sin\phi\, \frac{\partial}{\partial_\phi}\quad ,\\
v_{[2]}&=&\sin\phi\, \frac{\partial}{\partial\theta}+\cot\theta\cos\phi\, \frac{\partial}{\partial\phi}\quad .
\ea

Here we will not try to obtain these vector fields from the gauge transformation parameters. Instead, we will just list the gauge parameters corresponding to these vector fields. Requiring that the Lie derivative of $A^i$ with respect to these vector fields equals to a gauge transformation gives the following four corresponding Killing parameters
\ba
\xi^i_{[t]}&=&(0,0,0),\\
\xi^i_{[\phi]}&=&(0,0,0),\\
\xi^i_{[1]}&=&(-\csc\theta\sin\phi,0,0),\\
\xi^i_{[2]}&=&(\csc\theta\cos\phi,0,0).
\ea
Two of these are zero, as the Lie derivative of the connection is simply zero for the corresponding vector fields. However, the shifted gauge transformation parameters (\ref{shifted}) are not zero and work out to be given by
\ba
\tilde\xi^i_{[t]}&=&(-a,0,0),\\
\tilde\xi^i_{[\phi]}&=&(-\cos\theta,b\sin\theta,0),\\
\tilde\xi^i_{[1]}&=&(-\sin\theta\sin\phi,-b\cos\theta\sin\phi,-b\cos\phi),\\
\tilde\xi^i_{[2]}&=&(\sin\theta\cos\phi,b\cos\theta\cos\phi,-b\sin\phi).
\ea
We know from general considerations that these gauge parameters satisfy the equations (\ref{eqn-proj}), and the Killing vector fields (\ref{KVF-spherical}) can be recovered from them. Interestingly, we note that for $b=1$ the set of last three gauge transformation parameters forms an orthonormal triad in $\R^3$.

\end{document}